# Title: Monolayer Excitonic Laser

Yu Ye[1,2,†], Zi Jing Wong[1,2,†], Xiufang Lu[3], Hanyu Zhu[1], Xianhui Chen[3], Yuan Wang[1,2], Xiang Zhang[1,2,*]

[1] *NSF Nanoscale Science and Engineering Center, University of California, Berkeley, CA 94720, USA.*

[2] *Materials Sciences Division, Lawrence Berkeley National Laboratory, Berkeley, CA 94720, USA.*

[3] *Hefei National Laboratory for Physical Science at Microscale and Department of Physics, University of Science and Technology of China, Hefei, Anhui 230026, China.*

[†] *These authors contributed equally to this work.*

[*] *Correspondence and requests for materials should be addressed to X. Z. (email: xiang@berkeley.edu)*

**Abstract:**

Recently, two-dimensional (2D) materials have opened a new paradigm for fundamental physics explorations and device applications. Unlike gapless graphene, monolayer transition metal dichalcogenide (TMDC) has new optical functionalities for next generation ultra-compact electronic and opto-electronic devices. When TMDC crystals are thinned down to monolayers, they undergo an indirect to direct bandgap transition[1, 2], making it an outstanding 2D semiconductor[1-5]. Unique electron valley degree of freedom[6-9], strong light matter interactions[10] and excitonic effects were observed[11-13]. Enhancement of spontaneous emission has been reported on TMDC monolayers integrated with photonic crystal[14, 15] and distributed Bragg reflector microcavities[16, 17]. However, the coherent light emission from 2D monolayer TMDC has not been demonstrated, mainly due to that an atomic membrane has limited material gain volume and is lack of optical mode confinement. Here, we report the first realization of 2D excitonic laser by embedding monolayer tungsten disulfide ($WS_2$) in a microdisk resonator. Using a whispering gallery mode (WGM) resonator with a high quality factor and optical confinement, we observed bright excitonic lasing in visible wavelength. The $Si_3N_4$/$WS_2$/HSQ sandwich configuration provides a strong feedback and mode overlap with monolayer gain. This demonstration of 2D excitonic laser marks a major step towards 2D on-chip optoelectronics for high performance optical communication and computing applications.

As a direct bandgap semiconductor, Transition metal dichalcogenide (TMDC) monolayers have attracted increasing attention in electronics and optoelectronics applications due to their strong light emission accompanied with the unique access to spin and valley degrees of freedom[6-8, 18]. These properties arise from the quantum confinement and crystal symmetry effect on the band structure, as the material is thinned down to monolayer configuration. Excellent on-off ratio transistors[19], valley-Hall device[9], large exciton binding energy[11-13], light-emitting diodes[3-5], superconductivity[20], sensors[21], and piezoelectricity[22, 23] have been reported. Yet, the coherent light emission or lasing, which is the essential ingredient towards the realization of on-chip photonics applications, has not been realized. The design and fabrication of microcavities is crucial for the 2D laser, which requires high optical mode confinement factor and high quality factor. Here, we demonstrated 2D excitonic laser using a monolayer tungsten disulfide ($WS_2$) coupled with microdisk resonator, with high quantum yield, small footprint and low power consumption.

TMDCs, for example $WS_2$ (Fig. 1a), evolve from indirect to direct bandgap semiconductors as the number of layers is reduced from bulk to monolayer, with sizable bandgaps around 2.0 eV in the visible range. The direct bandgaps sit at the *K* and *K'* valleys (Fig. 1b), two non-equivalent momentum valleys in the reciprocal space of a monolayer $WS_2$ protected from its broken inversion symmetry, providing a rich valley-contrasting physics[6-9, 18]. The transition between valence band and conduction band edge is excitonic in nature in such a monolayer system. The strong excitonic features, including the neutral and redshifted charged excitons are observed and studied[24]. The exciton in 2D TMDCs not only governs the emissions properties, but also allows long-lived population inversion required to attain optical gain and possible stimulated emissions. Although Purcell enhancement of spontaneous emission has been achieved in photonic crystal and distributed Bragg reflector microcavities[14-17], the coherent light emission or lasing from 2D semiconducting TMDC has not been demonstrated due to the limited material gain volume of the atomic membrane and lack of optical confinement and feedback.

Here we reported a monolayer excitonic laser in a microdisk resonator (Fig. 1c). By integrating the monolayer $WS_2$ in a strong feedback photonic cavity, the build-up of stimulation emission can eventually exceed the lasing threshold. Microdisks feature low-loss, high-quality whispering gallery modes (WGMs) that offer the potential for ultralow-threshold lasing[25]. Embedding the monolayer in between two dielectric layers ($Si_3N_4/WS_2$/HSQ) attains strong optical confinement and leads to a larger modal gain, particularly crucial for an atomically thin monolayer gain medium. The scanning electron microscope (SEM) image (Fig. 1d) of the undercut $Si_3N_4/WS_2$/HSQ microdisk shows smooth sidewall roughness, essential to attain high cavity quality factor (Q). The diameter of the HSQ layer is slightly smaller than that of $Si_3N_4$ layer due to the finite etching selectivity between the HSQ and the undercut silicon layer during the xenon difluoride etching.

The cavity resonance is designed to overlap with the gain spectrum of monolayer $WS_2$, with electric field polarized in the plane of TMDC monolayer to efficiently couple with the in-plane dipoles of the excitons[26]. For a $Si_3N_4$/HSQ microdisk structure of diameter 3.3 um, we thus expect a strong transverse electric (TE) -polarized whispering gallery mode to show up at wavelength of around 612 nm. The small diameter of the microdisk ensures other resonance modes are widely separated in wavelength to avoid mode competition. The reducing number of modes will also increase the spontaneous emission and contribute to the improvement in the lasing threshold. The electric field distribution of the $TE_{1,\,24}$ resonance (radial mode number $l$=1, azimuthal mode number $m$=24), top view (Fig. 2a) and cross sectional view (Fig. 2b) are shown, and the resonant wavelength matches with the dominant peak of the measured lasing spectrum of a $WS_2$ monolayer embedded 3.3 μm diameter microdisk at 612.2 nm (Fig. 2c), with a measured quality factor $Q=\lambda/\Delta\lambda$ of about 2,604. The sandwiched configuration provides multiple advantages: (1) enhanced optical mode overlap, (2) material protection. It is important to note that the optical confinement factor of our sandwich $Si_3N_4$/$WS_2$/HSQ structure is about 30% higher than the case of directly transferring the monolayer to the top of a pre-built microdisk. The enhanced confinement factor is one order larger than the low-threshold quantum-dots microdisk lasers[27]. It is worth mentioning that the lasing performance of our 2D laser does not decay even after several months, as our sandwich $Si_3N_4$/$WS_2$/HSQ structure essentially protects the monolayer from direct exposure to air, which is known to degrade its luminescence property[28]. In addition, the whispering gallery resonance feature of our structure is further verified by the three additional peaks in the PL spectrum, i.e. at 633.7 nm, 657.6 nm and 683.7 nm, corresponding to the $TE_{1, 23}$, $TE_{1, 22}$ and $TE_{1, 21}$ modes in simulation. Such modes are only observed on the low-energy side of the spectrum, as high-energy photons are more likely to be reabsorbed before coupling to cavity resonance[25].

To study the emission characteristics of the monolayer $WS_2$ microdisk laser, we optically pumped the device with an ultrafast laser (190 fs pulse duration, 80 MHz repetition rate) at 473 nm at 10 K. The device has a diameter of 3.3 μm, and the emission spectra were captured by a 50× objective of 0.55 NA and collected as a function of pump intensity (Fig. 3a). At low pump intensity, the emission is broad. For pump intensity between 3.14 MW cm$^{-2}$ and 22.4 MW cm$^{-2}$, we observe a shoulder appearing at 612.2 nm, amplified by the optical feedback in the microdisk cavity. Above pump energy of 22.4 MW cm$^{-2}$, the peak increases sharply in intensity showing a clear evidence of lasing. The 2D nature of TMDC monolayer modifies the density of states to a step-like density of states, which realizes high gain and narrows the gain spectrum compared to those of bulk materials. This modification of the density of states results in several improvements in laser characteristics such as lower threshold, higher modulation bandwidth, and smaller emission linewidth, as were first predicted theoretically and then demonstrated experimentally in quantum-well lasers[29]. The 10 K PL spectrum at 65.7 MW cm$^{-2}$ pump power (Fig. 3b) is well fitted with bi-Lorentzian curves, where the peak intensity of both the broad monolayer WS$_2$ PL spectrum background and narrow cavity emission are extracted. In Fig. 3c, pump intensity dependence of the PL emission (*L-L* curves) is plotted for both the cavity and the background PL emission. Below the lasing threshold, the cavity emission increases linearly with excitation energy. Above the threshold, a distinct kink is observed in the *L-L* curve, with a super-linear increase in emission output. In contrast, the monolayer WS$_2$ background PL emission maintains a linear dependence on all pump energy. It is worth noting that the longer-wavelength WGMs discussed in Fig. 2 does not show lasing behaviors, due to the insufficient optical gain at wavelengths off the resonance of the exctonic transition in monolayer WS$_2$.

The internal luminescence quantum yield, defined by the fractions of absorbed photons that emit radiatively, is indeed one of the key elements to build a laser. High quantum yield is crucial to attain large optical gain, and is desirable for lowering the threshold pump power. Mak *et al* found a factor of 10$^4$ increasing in photoluminescence quantum yield from bulk MoS$_2$ to monolayer[4]. However, the quantum yield of monolayer MoS$_2$ measured thus far is only 4.3×10$^{-3}$, which is significantly lower than the near-unity values of many direct bandgap semiconductors such as GaAs, CdS. Here we experimentally measured the PL of MoS$_2$, WS$_2$ and WSe$_2$ monolayers, and quantified their respective quantum yields to determine the best monolayer TMDC material for lasing applications in our lab. WS$_2$ and WSe$_2$ monolayers are exfoliated from synthetic crystals grown by the chemical vapor transport method, while the MoS$_2$ monolayer is exfoliated from a natural crystal (SPI supplies). The monolayer quantum yield is extracted by measuring the absorption (in addition to PL) and calibrated using standard rhodamine 6G samples as the control[1, 30]. At room-temperature, it is found that the extracted quantum yield of the WS$_2$ monolayer has similar value to that of the WSe$_2$ monolayer, and is close to two orders larger than that of the MoS$_2$ monolayer, as shown in the left inset of Fig. 4. The observed difference in the three TMDC monolayer materials is likely due to the unintentional doping during the synthetic/natural growth process of the crystals. In addition, we study the temperature-dependence of the photoluminescence and quantum yield. Figure 4 shows the PL evolution of a monolayer WS$_2$ from room temperature 300 K down to cryogenic temperature 10 K. The two peaks observed at around 2.06 eV and 2.02 eV originate from the neutral exciton and charged exciton emission (corresponding to excitonic transition in Fig. 1b). The charged excitons are formed when photogenerated excitons bind to free electrons, indicating an unintentional doping of the monolayer[19]. No defect related emission is observed. As the temperature decreases, both the neutral and charged exciton peaks become narrower and shift to higher energy. More importantly, the amplitudes of both emissions gradually increase when the sample is cooled down from 300 K to 10 K. The extracted quantum yield of monolayer WS$_2$ at 10 K increases more than one order compared to that of 300 K (right inset of Fig.4). Deteriorating mechanisms, such as Auger recombination and free carrier absorption are strong at room temperature, thereby limiting the quantum yield[31], whereas low temperature environment prolongs the lifetime of Auger recombination and suppresses the non-radiative recombination, resulting in a much higher quantum yield[32]. Similarly, monolayer MoS$_2$ and WSe$_2$ show a significantly enhanced quantum yield at 10 K, which, however, is accompanied by defect emission below the band edge[33, 34], which is negligible for monolayer WS$_2$. In our measurement, the monolayer WS$_2$ has the largest quantum yield at 10 K (about 6%), which is around 5 times and two orders larger than those of monolayer WSe$_2$ and MoS$_2$ respectively. that is why we chose WS$_2$ monolayer in our lab for optical gain and laser.

The high quantum yield and efficient suppression of nonradiative recombination of monolayer WS$_2$ at low temperature allows the onset of stimulated emission at relatively low pump intensity. We use $\Gamma g_{th} = (2\pi\tilde{n}/\lambda)(1/Q)$ to estimate the threshold gain required to sustain the lasing modes, where $\Gamma$ is the mode confinement factor, $g_{th}$ is the threshold gain, $\tilde{n}$ is the mode effective refractive index, $\lambda$ is the emission wavelength and $Q$ is the overall quality factor. The calculated threshold gain for our 3.3 μm microdisk with $Q$ of 2,604 is 67,000 cm$^{-1}$. Within the range of pump power used in this experiment, no PL intensity saturation was observed, indicating that the exciton-exciton annihilation is negligibly small at 10 K.

In summary, we demonstrated visible 2D excitonic laser emission using monolayer WS$_2$ as a novel gain medium. The ability to generate coherent light radiation in these atomically thin semiconducting membranes paves the way for 2D coherent control and layered-material based optoelectronics. Selective excitation electron population in one of the valley can further lead to valley-polarized lasing, a key element for future valley optoeletronics.

Figure 1

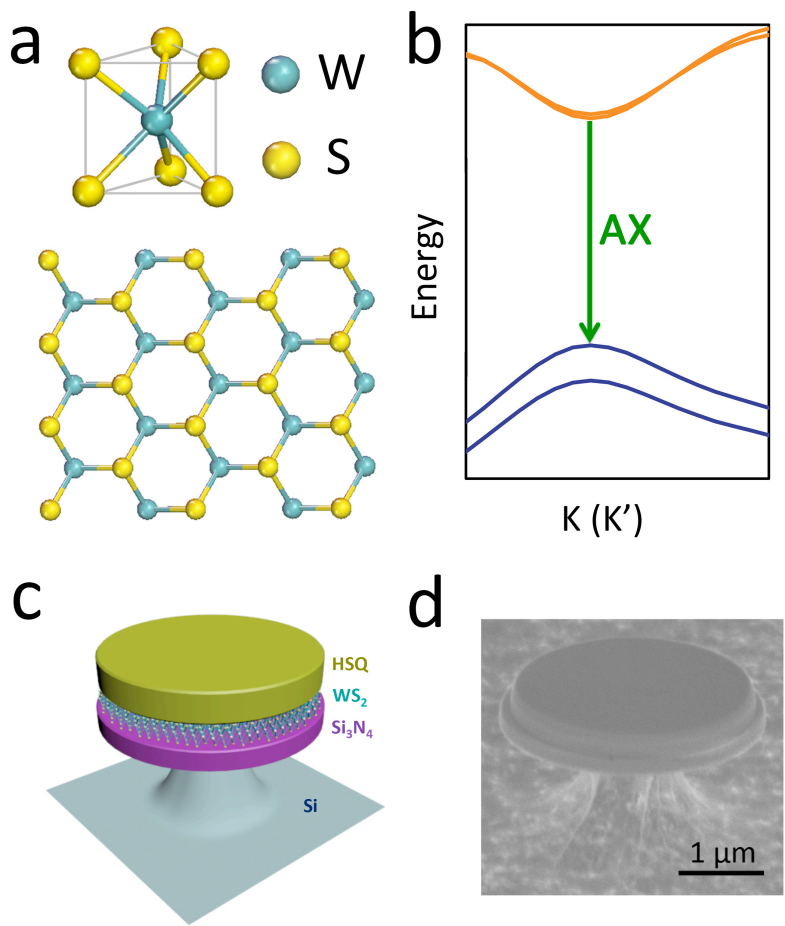

**Figure 1. Monolayer exitonic laser.** a, A single layer of $WS_2$ consists of S-W-S stacking with a total thickness of 0.65 nm. From the top view each unit cell consists of two sulfur atoms occupying the same site in the hexagonal lattice while the tungsten atom residing at the opposite site. b, Band structure at *K* (*K'*) point shows the direct band *A* exciton transition, and valence band splitting due to spin-orbit coupling. c, Schematic image of monolayer $WS_2$ microdisk laser. The sandwich structure of $Si_3N_4$/$WS_2$/HSQ ensures a higher confinement factor and leads to a larger modal gain. d, SEM of an undercut $Si_3N_4$/$WS_2$/HSQ microdisk, showing the smooth sidewall essential to attain high cavity quality factor.

**Figure 2**

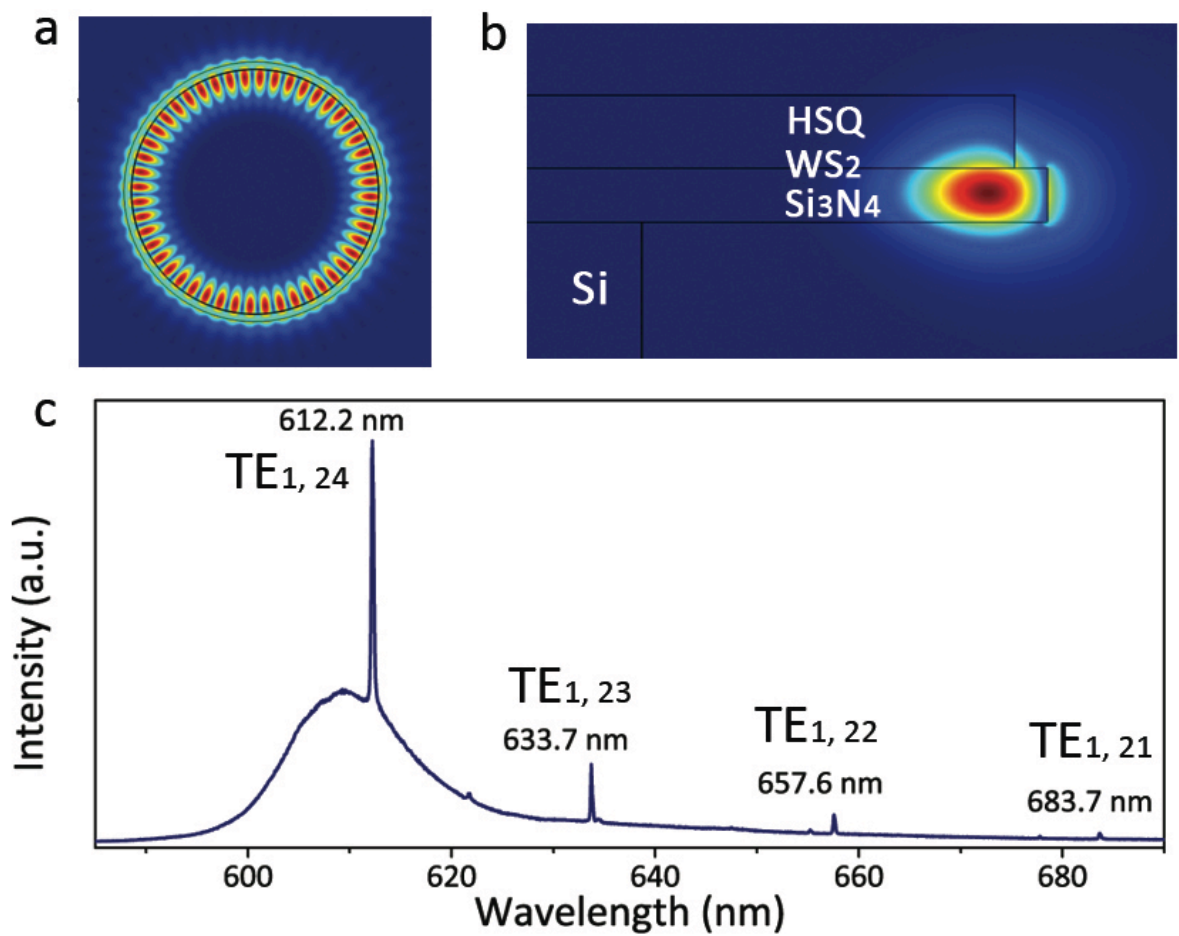

**Figure 2. The lasing mode of monolayer excitonic laser.** a, The top view of simulated electric field distribution of the $TE_{1,24}$ resonance. b, The cross sectional view of the calculated electric field distribution of $TE_{1,24}$ resonance. The $Si_3N_4/WS_2/HSQ$ sandwich structure enhances the optical mode overlap with monolayer $WS_2$. Red (blue) color corresponds to the maximum (minimum) of the field density. c, Experimental photoluminescence (PL) spectrum taken at 10 K when the pump intensity is above lasing threshold, showing whispering galley modes at 612.2 nm, 633.7 nm, 657.6 nm and 683.7 nm.

**Figure 3**

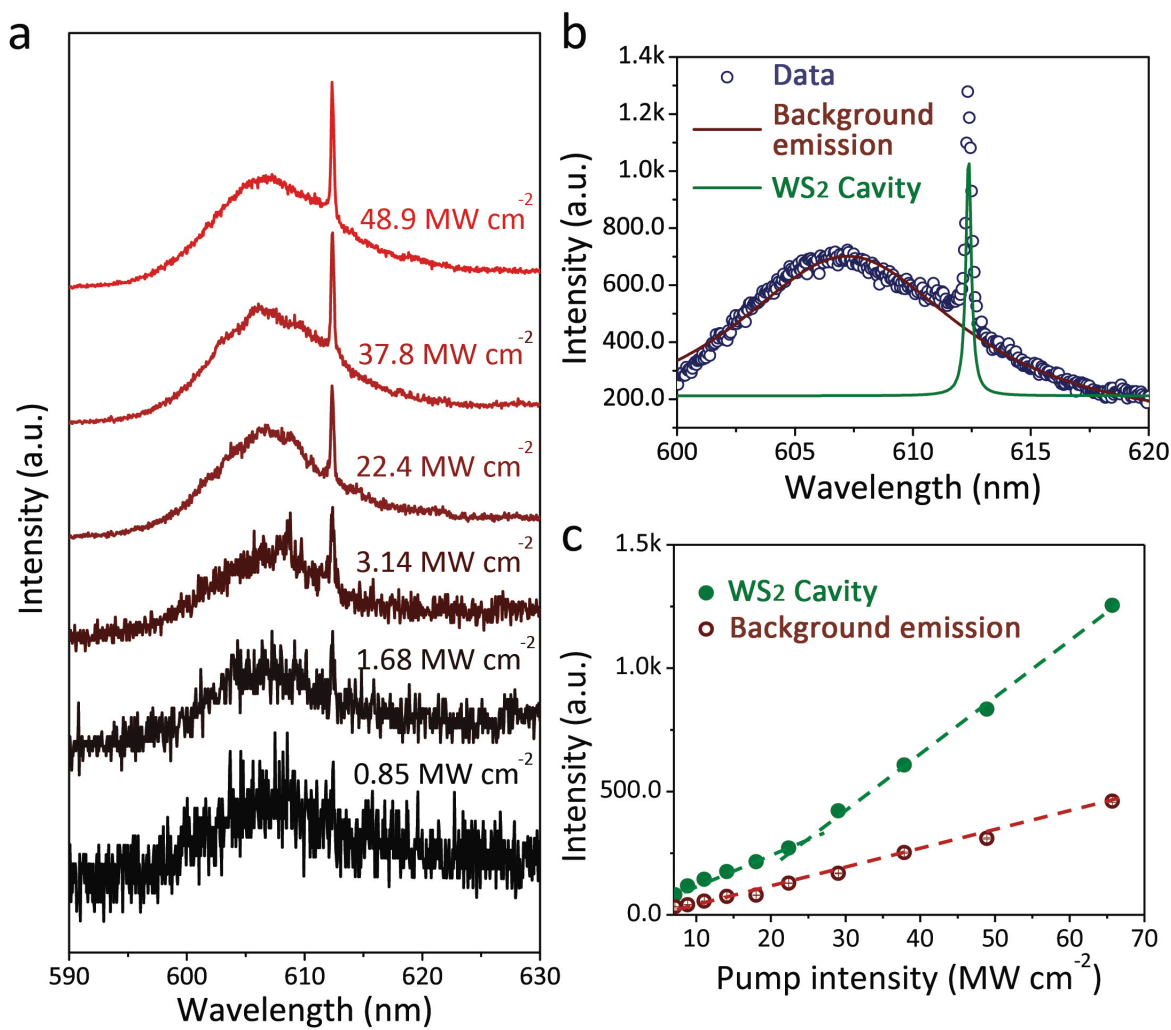

**Figure 3. Monolayer WS$_2$ laser characteristics.** a, Steady-state PL emission spectra from a 3.3 μm microdisk photoexcited using 473 nm, 190 fs and 80 MHz pump pulses with increasing pump intensity, normalized to pump intensity, illustrating the transition from spontaneous emission to stimulated emission and lasing. b, PL spectrum under pump intensity of 65.7 MW cm$^{-2}$. The dots are the measured data, the brown line is the fitting to the monolayer WS$_2$ PL background emission (mainly from the center of the microdisk), and green line is the fitting to the WS$_2$ cavity emission. c, Monolayer WS$_2$ PL background and cavity emissions as a function of pump intensity. The dashed lines represent the linear fits to the experimental data. The WS$_2$ PL background emission shows linear dependence of the pump intensity, while the green dash lines (cavity emission) shows a kink indicating the onset of superlinear emission and lasing operation.

**Figure 4**

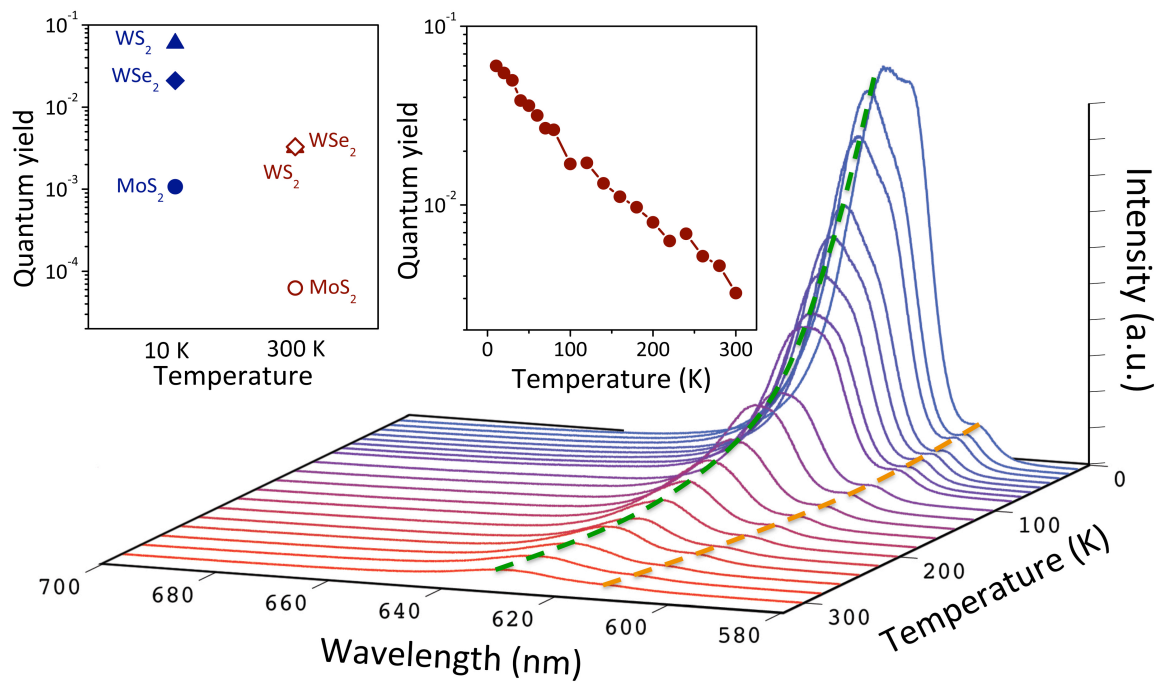

**Figure 4. Quantum yield measurement for different types of monolayer TMDCs.**
Evolution of the emission spectra of the monolayer $WS_2$ when temperature drops from 300 K to 10 K. The A exciton emission intensity monotonically increases with the cool down of the sample. Left inset: the extracted PL quantum yield of monolayer $MoS_2$, $WS_2$ and $WSe_2$ at 300 K and 10 K. The monolayer $WS_2$ at 10 K has highest quantum yield (about 6%), which is five times and two orders larger than those of $WSe_2$ and $MoS_2$. Right inset: temperature-dependence of the quantum yield of a monolayer $WS_2$. The quantum yield of a monolayer $WS_2$ increases more than one order of magnitude when cooled down from 300 K to 10 K.